\documentclass[fleqn,usenatbib]{mnras}

\usepackage{graphicx}	
\usepackage{amsmath}	
\usepackage{amssymb}	
\usepackage{multicol}        
\usepackage{bm}		
\usepackage{pdflscape}	

\begin{document}

\title[Cosmological parameter from transversal BAO]{Cosmological parameter analyses using transversal BAO data}

\author[R. C. Nunes, S. K. Yadav, J. F. Jesus, A. Bernui]{
Rafael C. Nunes,$^{1}$\thanks{E-mail: rafadcnunes@gmail.com}
Santosh K. Yadav,$^{2}$\thanks{E-mail: sky91bbaulko@gmail.com}
J. F. Jesus,$^{3,4}$\thanks{E-mail: jf.jesus@unesp.br}
Armando Bernui$^{5}$\thanks{E-mail: bernui@on.br}
\\
$^{1}$Divis\~ao de Astrof\'isica, Instituto Nacional de Pesquisas Espaciais, Avenida dos Astronautas 1758, S\~ao Jos\'e dos Campos, 12227-010, SP, Brazil\\
$^{2}$Department of Mathematics, BITS Pilani, Pilani Campus, Rajasthan-333031, India\\
$^{3}$UNESP - C\^ampus Experimental de Itapeva, Itapeva, SP, Brazil\\
$^{4}$UNESP - Faculdade de Engenharia de Guaratinguet\'a, 
Guaratinguet\'a, SP, Brazil\\
$^{5}$ Observat\'orio Nacional, 
Rua General Jos\'e Cristino 77, S\~ao Crist\'ov\~ao, 20921-400, Rio de Janeiro, RJ, Brazil
}

\date{Accepted XXX. Received YYY; in original form ZZZ}

\pubyear{2020}

\label{firstpage}
\pagerange{\pageref{firstpage}--\pageref{lastpage}}
\maketitle

\begin{abstract}
We investigate observational constraints on cosmological parameters combining
15 measurements of the transversal BAO scale (obtained free of any fiducial cosmology) with Planck-CMB data to explore the parametric space of some cosmological models. We investigate how much Planck + transversal BAO data can constraint the minimum $\Lambda$CDM model,  and extensions, including neutrinos mass scale $M_{\nu}$, and the possibility for a dynamical dark energy (DE) scenario. Assuming the $\Lambda$CDM cosmology, we find $H_0 = 69.23 \pm 0.50$ km s${}^{-1}$ Mpc${}^{-1}$, $M_{\nu} < 0.11$~eV and $r_{\rm drag} = 147.59 \pm 0.26$ Mpc (the sound horizon at drag epoch) 
from Planck + transversal BAO data. When assuming a dynamical DE cosmology, we find that the inclusion of the BAO data can indeed break the degeneracy of the DE free parameters, improving the constraints on the full parameter space significantly. We note that the model is compatible with local measurements of $H_0$ and there is no tension on $H_0$ estimates in this dynamical DE context. Also, we discuss constraints and consequences from a joint analysis with the local $H_0$ measurement from SH0ES. Finally, we perform a model-independent analysis for the deceleration parameter, $q(z)$, using only information from transversal BAO data.
\end{abstract}


\section{Introduction}
\label{sec:level1}

Nowadays are stimulating for cosmology, with a plethora of dark energy models 
competing to describe the current high-quality cosmological observations~\citep{SDSS,PLA}.  The combination of data from cosmological probes like standard candles, from type Ia supernovae data~\citep{snia}, standard rulers, from Baryon Acoustic Oscillations (BAO) 
data~\citep{Eisenstein05,Aubourg,Ata}, together with measurements of the cosmic microwave background (CMB) radiation~\citep{wmap9,planck_VI} have produced precise parameters constraints of the concordance cosmological model, the spatially flat $\Lambda$CDM~\citep{planck_VI,sdss1,des1},  strongly restricting alternative scenarios (see, e.g.,~\cite{M01,M02,M03,M04,M05,M06,M07,M08,M09,M10}). 

To investigate cosmological models or parameters using observational data one usually performs a likelihood approach, which is done under basic assumptions. 
At the end, some hypothesis can be changed and the possible dependence of the results on such modification is studied. This is the case of the analyses of the CMB temperature fluctuations measurements, based on a power-law spectrum of adiabatic scalar perturbations, done by the Planck collaboration using a combination of temperature, polarization, and lensing CMB data. They found~\citep{planck_VI} that the best-fit values of the spatially flat six parameter (i.e., 
$\Omega_{b}\,h^2$, $\Omega_{c}\,h^2$, $\theta_{\star}$, $\tau$, $A_s$, $n_s$) $\Lambda$CDM model (here termed {\em minimum} $\Lambda$CDM) provides a good consistency with the data, with no indications for a preference on extensions of this basic set of parameters and hypotheses. Assuming this {\em minimum} $\Lambda$CDM cosmology, the Planck collaboration also found derived model-dependent parameters: $H_0, \Omega_m, \mbox{\rm and}\, \sigma_8$~\citep{planck_VI}. 
Additionally, combining CMB data with BAO measurements, they found~\citep{planck_VI} that the neutrino mass is tightly constrained to $M_{\nu} < 0.12$ eV, and that the effective extra relativistic  degrees of freedom agrees with the prediction of the Standard Model $N_{\rm eff} = 3.046$. 

One cannot expect that the whole parameter space of the model will be well-behaved when describing the set of distinct cosmological probes used in several combined analysis. This is the case with the CMB lensing amplitude and the parameters related to, in fact, it was reported that the CMB spectra prefer higher lensing amplitudes than predicted in the {\em minimum} $\Lambda$CDM at over $2\sigma$~\cite{planck_VI}. Another recognized tension is that one reported for the Hubble parameter $H_0$: while local SNIa observations measure $H_0=74.03 \pm 1.42$~\citep{Riess2019}, the Planck collaboration report $H_0=67.4 \pm 0.5$ km s$^{-1}$Mpc$^{-1}$~\cite{planck_VI}. And the question is how much dependent on model hypotheses are these  results ~\citep{Sutherland,Heavens} or do they reflect just measurements with underestimated systematics ~\cite{Shanks}.
For examples, of how systematics can  influence data 
analyses see, e.g.,~\cite{BTV,BTV2,BFW,Bengaly,Novaes14,Novaes15, Marques}.

Current data analyses combine diverse cosmological probes to break degeneracy between 
cosmological parameters using, for instance, the data from Type IA supernova data or from 
Baryon Acoustic Oscillations (BAO). 
Regarding the use of the BAO data, an interesting issue refers to a possible bias in the 
standard BAO analysis due to the assumption of a fiducial cosmological model needed to 
calculate the fiducial comoving-coordinates, however, recent studies conclude that such 
assumption does not introduce any bias (see, e.g.,~\citet{Aubourg,Vargas,Carter}). 
From the other side, the transversal BAO analysis does not need to assume a fiducial 
cosmology, and was also used for clustering analysis (see, e.g.,~\citet{Sanchez11,Carnero,Gabriela1}), although the precision of the angular BAO scale measurements with this 
methodology is not competitive with the standard one. 
Nevertheless, transversal BAO analyses complement standard BAO analyses and are interesting because: 
(a) transversal BAO studies, as described by  \citet{Sanchez11,Carnero} 
are weakly dependent on the assumption of a cosmological model, 
this allows a comparison with the results obtained with the standard BAO approach (comparisons are welcome 
in natural sciences, just as a validation or confirmation of results obtained by a different team using a distinct  methodology); 
(b) since the transversal BAO only data needs angular distances between pairs to calculate the 2-point angular correlation function, these 
analysis (and their results) are independent of the 3D curvature parameter, $\Omega_k$, while in the standard BAO analysis this  parameter has to be assumed (notice that the curvature parameter is being subject of a recent controversy \cite{ValentinoK1,ValentinoK2}. 
(c) the methodology described by \citet{Sanchez11} is a simple  approach that lets to study the evolution of the Universe if one  performs a tomographic study analyzing data in contiguous thin 
redshift shells, which are disjoint and separate in more than the uncertainties of the redshift measurements, this avoids 
the correlation between shells turning the measurements of  $D_A(z)/r_{\rm drag}$ at several redshifts statistically  independent. See \cite{baot1,baot2,baot3,baot4,baot5,baot6,baot7} for others recent discussions on the cosmological constraints investigations under the perspective of the BAO measurements.

This work is organized as follows. 
In the next section, we provide an overview of contemporary cosmological parameters 
analyses, while in section~\ref{data} we present our data set and the statistical methodology 
adopted. 
In section~\ref{results} we present the results and the related discussion of our analyses. 
In section~\ref{qz_section} we perform a model-independent analysis on the $q(z)$ function 
from the transversal BAO data and, in section~\ref{final}, we summarize the findings of our 
analyses and express future perspectives.

\section{Cosmological parameter analyses}
\label{section2}

According to the above concerns, we perform cosmological parameter analyses combining CMB data with a set of 15 measurements of the transversal BAO scale measurements, 
obtained according to the model-independent  approach of \citet{Sanchez11}, to explore via 
Monte Carlo Markov  chains the parametric space of some cosmological models.
Moreover, the literature shows the efforts to measure the sound horizon 
$r_{\rm drag}$ in a weakly model-dependent approach (see, e.g.,~\citet{Sutherland,Heavens}). 
Following this objective, we shall combine data from Planck 2018 CMB data together with BAO, 
but using only the above mentioned 15 measurements of transversal 
BAO data $D_{\!A}(z;r_{\rm drag})$. 
Complementing these analyses, we add the local Hubble parameter 
measurement~\citep{Riess2019} as a prior allowing a still more precise determination of 
$r_{\rm drag}$ at low redshift, in specific cases where the $H_0$ tension is not present. 

On the other hand, the neutrinos play a crucial role in the dynamics of our Universe, by inferring direct changes in the clustering of  structures and, consequently, in the determination of cosmological parameters (see an incomplete list of works that investigates neutrino  features~\citep{neutrinos01,neutrinos02,neutrinos03,neutrinos05,neutrinos06,neutrinos07,neutrinos11} and references therein). 
The standard parameters that characterize these effects are the effective number of neutrino species $N_{\rm eff}$ and the total neutrino mass scale $M_{\nu}$. 
We refer to~\citep{planck_VI, neutrinos11} for recent constraints on these parameters. 
In principle, both quantities $N_{\rm eff}$ and $M_{\nu}$ are model dependent, and hence, 
different cosmological scenarios may bound these parameters in different ways. 
In our analysis, we consider that recent measurements of the transversal BAO signature can
help to a better constraint of the neutrinos mass scale.
It will be the first time that such data sets will be used to investigate these properties of neutrinos. 

According to this plan, we combine for the first time a compilation of  these angular BAO measurements, with the following  aims: \\

(i) To perform combined analyses that explores the parameters space of some dark energy (DE) models and, additionally, to obtain a precise estimate of $r_{\rm drag}$ from each model. This is due to the fact that the inclusion of transversal BAO data will improve estimates on matter density and Hubble constant, which, in return, will also improve the bounds on $r_{\rm drag}$.\\

(ii) To derive new bounds on the neutrino mass scale $M_{\nu}$ within $\Lambda$CDM and extended models, considering the possibility for some dynamical dark energy scenarios, robustly using transversal BAO data. \\

(iii) An independent model analysis to reconstruct the deceleration parameter $q(z)$ from our compilation of transverse BAO data.\\

The cosmological model analyses mentioned above in the points (i) and (ii) encompass two scenarios: the current concordance model, i.e., the spatially flat $\Lambda$CDM, and the simplest natural extension to the $\Lambda$CDM model, that is, a dynamical DE model, $w_0\mbox{-}w_a$CDM considering the Chevallier-Polarski-Linder (CPL) parameterization~\citep{cpl1,cpl2}, where the EoS is characterized by $w(a) = w_0 \,+\, w_a (1 - a)$, where $a$ is the scale factor in a Friedmann-Robertson-Walker cosmology. 

Our main results show that the joint analyses of CMB data and transversal BAO measurements improves the parameter constraints, also in the case where we include the neutrino mass as an extra-parameter. Moreover, we also recover the best-fit values of the {\em minimum} $\Lambda$CDM model~\citep{planck_VI}, concerning the main baseline and derived cosmological parameters, with improved values when including BAO data.

\section{Data and Methodology}
\label{data} 

In what follows, we describe the observational data sets used in this work. \\

\noindent
\textbf{BAO}: First, we describe the set of 15 transversal BAO measurements, $\theta_{\mbox{\sc bao}}(z)$, 
obtained without assuming a fiducial cosmological model~\citep{Sanchez11,Carnero}. 
In fact, in a thin redshift bin with suitable number density of cosmic tracers --like quasars or 
galaxies--, one can perform the 2-point angular correlation function between pairs to find and 
measure the BAO angular scale $\theta_{\mbox{\sc bao}}(z)$, at that redshift (see, 
e.g.,~\citep{Gabriela1} and refs. therein). 
A BAO angular scale measurement gives the angular diameter distance $D_A$ at the redshift 
$z$ 
\begin{equation}
\label{eq_bao}
D_A(z;r_{\rm drag}) \,=\, \frac{ r_{\rm drag} }{ (1+z) \, \theta_{_{\mbox{\sc bao}}}(z) } \, , 
\end{equation}
provided that one has a robust estimate of $r_{\rm drag}$, the comoving sound horizon at the baryon drag epoch. 

Our set of 15 transversal BAO measurements~\citep{Gabriela1,Alcaniz,Gabriela2,Edilson1,Edilson3}, were obtained using public data releases (DR) of the Sloan Digital Sky 
Survey (SDSS), namely: DR7, DR10, DR11, DR12, DR12Q (quasars)~\citep{SDSS}. These data set is displayed in Table~\ref{table1}. 
It is important to notice that, due to the cosmological-independent methodology used to 
perform these transversal BAO measurements their errors are larger than the errors obtained 
using a fiducial cosmology approach. 
The reason for this fact is that, while in the former methodology the error is given by the 
measure of how large is the BAO bump, in the later approach the model-dependent best-fit 
of the BAO signal quantifies a smaller error. 
Typically, in the former methodology the error can be of the order of $\sim 10\%$, but in some 
cases it can arrive to 18\%, and in the later approach it is of the order of few 
percent~\citep{Sanchez11}.
\\

\begin{table} 
\centering
 \begin{tabular}{|c|c|c|c|}
\hline
\,\,\,\,\,\,\,\, $z$ \,\,\,\,\,\,\,\,& \,\,\,$\theta_{\mbox{\sc bao}}$ [deg]\,\,\,  
&  \,\,\,$\sigma_{\mbox{\sc bao}}$ [deg]\,\,\, & \,\,\, ref. \, \\
\hline
\,\,\,0.11\,\,\,  &  19.8  &  3.26  & ~\citet{Edilson3}  \\
0.235  &  9.06  &  0.23  & ~\citet{Alcaniz}  \\
0.365  &  6.33  &  0.22  & ~\citet{Alcaniz}  \\
0.45    &  4.77  &  0.17  & ~\citet{Gabriela1}  \\
0.47    &  5.02  &  0.25  & ~\citet{Gabriela1}  \\
0.49    &  4.99  &  0.21  & ~\citet{Gabriela1}  \\
0.51    &  4.81  &  0.17  & ~\citet{Gabriela1}  \\
0.53    &  4.29  &  0.30  & ~\citet{Gabriela1}  \\
0.55    &  4.25  &  0.25  & ~\citet{Gabriela1}  \\
0.57    &  4.59  &  0.36  & ~\citet{Gabriela2}  \\
0.59    &  4.39  &  0.33  & ~\citet{Gabriela2}  \\
0.61    & 3.85   &  0.31  & ~\citet{Gabriela2}  \\
0.63    & 3.90   &  0.43  & ~\citet{Gabriela2}  \\
0.65    & 3.55   &  0.16  & ~\citet{Gabriela2}  \\
2.225  & 1.77   &  0.31   & ~\citet{Edilson1}  \\
\hline
\end{tabular}
\caption{A compilation of angular BAO measurements from luminous red galaxies, blue galaxies, and quasars catalogs, from diverse releases of the Sloan Digital Sky Survey, all of them obtained following the approach of~\citet{Sanchez11}.} 
\label{table1}
\end{table}

\noindent
\textbf{CMB}: To break the degeneracy, when considering the full parametric space of the models under study in this work, we shall combine this BAO data 
set (see Table~\ref{table1}) together 
with the CMB data from the final release of the Planck collaboration (2018), including the likelihood TT+TE+EE+lowE+lensing ~\citep{planck_VI,Planck2018_2,Planck2018_3}. Here we
just refer this data set as Planck.
\\

\noindent
\textbf{R19}: In some analyses, we shall consider the recently measured new local value of the Hubble constant by the Hubble Space Telescope (HST): $H_0=74.03 \pm 1.42$ km s${}^{-1}$ Mpc${}^{-1}$ as reported in~\citep{Riess2019}. 
This value of the Hubble constant is in tension, at 4.4$\sigma$, with the Planck 2018 cosmological 
parameters calculation within the minimum $\Lambda$CDM model~\cite{planck_VI}. 
We refer to this datum as R19. 
\\

Regarding the theoretical framework, let us consider two scenarios in our analyses. The first scenario, termed $\Lambda$CDM + $M_{\nu}$, described from the baseline
\begin{eqnarray}
\mathcal{P} \equiv \Bigl\{ \omega_{b}, \omega_{_{\rm CDM}}, 100\theta_{*}, \tau_{\rm reio}, n_{s},%
\log[10^{10}A_{s}], M_{\nu} \Bigr\} \, ,
\label{eq:parameter_space1}
\end{eqnarray}
where the first six parameters corresponds to the minimum $\Lambda$CDM model: the baryon 
and the cold dark matter energy densities $\omega_{\rm b}$ and $\omega_{\rm cdm}$, the ratio 
between the sound horizon and the angular diameter distance at decoupling $100\theta_{*}$, 
the reionization optical depth $\tau_{reio}$, and the spectral index and the amplitude of the scalar 
primordial power spectrum $n_{s}$ and $A_{s}$, respectively. 

With respect to the neutrino properties, we impose a prior of $M_{\nu} > 0$, ignoring a possible 
lower limit from the neutrino oscillations experiments and assuming fixed three neutrinos species, 
that is, $N_{\rm eff} = 3.046$. 
For the purposes of obtaining bounds on neutrino mass from the cosmological data, the prior 
$M_{\nu} > 0$ is adequate. 

The second scenario, we will consider a dynamical DE model, where the EoS is given in terms of the 
CPL parametrization. Let us call this model by $w_0\mbox{-}w_a$CDM model. 
In this case, the parametric space is written as 
\begin{eqnarray}
\mathcal{P} \equiv\! \Bigl\{\omega_{b}, \omega_{_{\rm CDM}}, 100\theta_{*}, \tau_{\rm reio}, n_{s}, \log[10^{10}A_{s}], w_0, w_a, M_{\nu} \Bigr\},
\label{eq:parameter_space1}
\end{eqnarray}
where $w_0$ and $w_a$, are free parameters that characterize the dynamics of the EoS, where for $w_0 = -1$ and $w_a = 0$, we recovered the $\Lambda$CDM model.

We use the publicly available \texttt{CLASS}~\citep{class} and \texttt{MontePython}~\citep{monte} 
codes to analyze the free parameters of the models defined above. 
We used Metropolis Hastings algorithm with uniform priors on the full baseline parameters 
to obtain correlated Markov Chain Monte Carlo samples. 
We have ensured the convergence of the chains for all parameters according to the 
Gelman-Rubin criterium. During the statistical analyses, we consider the flat priors on all parameters, where the common baseline 
parameters in all scenarios is: $100\omega_{b} \in [0.8 \,,\, 2.4]$, 
$\omega_{cdm} \in [0.01\,,\, 0.99]$, $100 \theta_{*} \in [0.5\,,\,2]$, $\tau_{\rm reio} \in [0.01\,,\,0.8]$, 
$\log_{10}(10^{10}A_{s}) \in [2\,,\,4]$, $n_s \in [0.9 \,,\, 1.1]$, $w_0 \in [-3 \,,\, 0]$, $w_a \in [-3 \,,\, 3]$, 
and $M_{\nu} \in [0 \,,\, 1]$. The $\Lambda$CDM model is obtained by fixing $w_0=-1$ and $w_a=0$. We assume a spatially flat Universe in all analyzes performed in this work. As usual, the parameters $H_0, \sigma_8, r_s$, and $\Omega_m$  are derived parameters.
In what follows we discuss our results. 

\begin{table*} 
\begin{tabular} {c| c c | c c}  
\hline \hline  

     &   \multicolumn{2}{c|}{Planck} & \multicolumn{2}{c}{Planck + BAO}  \\ 
 \hline
  Parameter &  $\Lambda$CDM  & $\Lambda$CDM + $M_{\nu}$  &   $\Lambda$CDM  & $\Lambda$CDM + $M_{\nu}$  \\  
 \hline

 $10^{2}\omega_{\rm b }$ & $2.240^{+ 0.015+0.031}_{-0.015-0.029}$  &  $2.233^{+ 0.015+0.029}_{-0.015-0.029}$  &  
 
    $2.260^{+0.014+0.028}_{-0.014-0.027} $ 
 &  $2.258^{+ 0.015+0.027}_{-0.014-0.027}$   \\
 
 $\omega_{_{\rm CDM }}  $ & $0.1199^{+ 0.0012+0.0024}_{-0.0012-0.0024}$  &  $0.1206^{+ 0.0013+0.0025}_{-0.0013-0.0024}$  & 
 
  $0.1206^{+0.0013+0.0025}_{-0.0013-0.0024} $ 
 &  $0.1172^{+0.0011+0.0021}_{-0.0011-0.0021}$  \\

$100 \theta_{*} $ & $1.0419^{+0.0003+0.0006}_{-0.0003-0.0006}$ &  $1.0418^{+0.0003+0.0006}_{-0.0003-0.0006}$
&  $1.0421^{ +0.0003+0.0005}_{-0.0003-0.0006}$ &  $1.0421^{ +0.0003+0.0005}_{-0.0003-0.0006} $  \\

$\ln10^{10}A_{s }$  & $ 3.044^{+0.014+0.027}_{-0.014-0.028}   $ & $3.048^{+0.015+0.030}_{-0.015-0.029}  $  & $3.055^{+0.016+0.031}_{-0.016-0.030}   $ &  $3.061^{+ 0.016+0.031}_{-0.016-0.030} $ \\

$n_{s } $ & $0.965^{+0.004+0.008}_{-0.004-0.008} $  &  $0.963^{+0.004+0.008}_{-0.004-0.008}  $    &  $0.971^{+0.004+0.008}_{-0.004-0.008}    $ &  $0.971^{+0.004+0.008}_{-0.004-0.008}  $ \\

$\tau_{\rm reio }   $ &  $0.054^{+0.007+0.015}_{-0.007-0.014}   $ &$0.055^{+0.007+0.015}_{-0.008-0.014}  $    &$0.062^{+0.007+0.016}_{-0.008-0.015} $  & $0.064^{+ 0.008+0.016}_{-0.008-0.016}$  \\

 $M_{\nu}$   & $--$   &  $ < 0.34$    & $--$ & $< 0.11$    \\ 
 
$\Omega_{m }$  & $0.308^{+0.007+0.015}_{-0.007-0.014} $  &  $0.326^{+ 0.009+0.025}_{-0.013-0.022}   $ &  $0.292^{+ 0.006+0.012}_{-0.006-0.012}  $ &  $0.299^{+ 0.006+0.014}_{-0.006-0.013}  $   \\ 

  $H_{\rm 0}$ & $67.39^{+0.56+1.10}_{-0.56-1.10} $ &  $ 66.59^{+ 0.96+1.60}_{-0.67-1.80}$ &  $69.23^{+0.50+1.00}_{-0.50-0.97} $ & $68.58^{+0.54+1.00}_{-0.54-1.10} $ \\ 
 
  $\sigma_{8} $ & $0.82^{+0.006+0.012}_{-0.006-0.012}  $ &  $0.800^{+ 0.0150+0.022}_{-0.007-0.028} $  & $ 0.819^{+ 0.006+0.013}_{-0.006-0.012}  $  & $ 0.807^{+0.006+0.013}_{-0.006-0.013} $ \\ 
  
$r_{\rm drag}$   & $ 147.09^{+ 0.27+0.52}_{-0.27-0.51}$ &  $ 147.97^{+ 0.28+0.53}_{-0.28-0.56}$ & 
$ 147.59^{+ 0.25+0.48}_{-0.25-0.48}$& $ 147.59^{+ 0.26+0.51}_{-0.26-0.52}$   \\ 
  
  \hline
 
  \end{tabular}
  \caption{Constraints at 68\% and 95\% CL on free and some derived parameters under $\Lambda$CDM  model baseline from the considered data combinations. 
The parameter $H_0$ is measured in the units of km/s/Mpc, $r_{\rm drag}$ in Mpc, whereas 
$M_{\nu}$ is in the units of eV.} 
\label{LCDM_results}
\end{table*}


\begin{table*} 
\begin{tabular} {c| c c  | c c}  
\hline \hline  

     &   \multicolumn{2}{c|}{Planck} & \multicolumn{2}{c}{Planck + BAO}  \\ 
 \hline
  Parameter &  $w_0\mbox{-}w_a$CDM  & $w_0\mbox{-}w_a$CDM + $M_{\nu}$  &   $w_0\mbox{-}w_a$CDM  & $w_0\mbox{-}w_a$CDM + $M_{\nu}$  \\  
 \hline 

 $10^{2}\omega_{\rm b }$ & $2.245^{+ 0.015+0.028}_{-0.015-0.028}$  &  $2.237^{+ 0.015+0.030}_{-0.015-0.029}$  &  
 
    $2.243^{+0.013+0.029}_{-0.015-0.026} $ 
 &  $2.240^{+ 0.015+0.029}_{-0.015-0.029}$   \\ 
 
 $\omega_{_{\rm CDM }}  $ & $0.1192^{+ 0.0013+0.0028}_{-0.0013-0.0026}$  &  $0.1299^{+ 0.0013+0.0027}_{-0.0013-0.0025}$  & 
 
  $0.1192^{+0.0011+0.0022}_{-0.0011-0.0023} $ 
 &  $0.1198^{+0.0012+0.0023}_{-0.0012-0.0023}$  \\ 

$100 \theta_{*} $ & $1.0419^{+0.0003+0.0006}_{-0.0003-0.0006}$ &  $1.0419^{+0.0003+0.0006}_{-0.0003-0.0006}$
&  $1.0421^{ +0.0003+0.0005}_{-0.0003-0.0005}$ &  $1.0421^{ +0.0003+0.0005}_{-0.0003-0.0006} $  \\ 

$\ln10^{10}A_{s }$  & $ 3.038^{+0.015+0.029}_{-0.015-0.030}   $ & $3.043^{+0.015+0.030}_{-0.015-0.030}  $  & $3.037^{+0.015+0.031}_{-0.015-0.029}   $ &  $3.043^{+ 0.014+0.030}_{-0.014-0.027} $ \\ 

$n_{s } $ & $0.967^{+0.004+0.008}_{-0.004-0.008} $  &  $0.965^{+0.004+0.008}_{-0.004-0.009}  $    &  $0.967^{+0.004+0.008}_{-0.004-0.008}    $ &  $0.965^{+0.004+0.008}_{-0.004-0.008}  $ \\ 

$\tau_{\rm reio }   $ &  $0.052^{+0.007+0.014}_{-0.007-0.015}   $ &$0.054^{+0.007+0.015}_{-0.007-0.014}  $    &$0.052^{+0.008+0.015}_{-0.008-0.015} $  & $0.054^{+ 0.007+0.016}_{-0.007-0.014}$  
     \\ 

$w_{\rm 0}$ &  $-1.28^{+0.42+0.71}_{-0.68-0.72}   $ &$-1.51^{+0.56+0.62}_{-0.48-0.50}  $    &$-0.92^{+0.29+0.39}_{-0.14-0.52} $  & $-1.04^{+ 0.30+0.42}_{-0.15-0.53}$  
     \\ 

$w_{\rm a}$ &  $-0.70^{+0.77+1.50}_{-1.30-1.30}   $ &$-0.58^{+0.86+1.30}_{-0.86-1.40}  $    &$-1.11^{+0.28+1.60}_{-0.86-0.94} $  & $-1.11^{+ 0.28+1.60}_{-0.88-0.93}$  \\ 

$ M_{\nu}$   & $--$   &  $ < 0.38$    & $--$ & $< 0.33$    \\ 
 
$\Omega_{m}$  & $0.222^{+0.073+0.120}_{-0.073-0.120} $  &  $0.205^{+ 0.024+0.150}_{-0.081-0.092}   $ &  $0.259^{+ 0.021+0.035}_{-0.017-0.040}  $ &  $0.253^{+ 0.021+0.038}_{-0.018-0.040}  $   \\ 

  $H_{\rm 0}$ & $83.0^{+10.0+30.0}_{-20.0-20.0} $ &  $ 87.0^{+ 10.0+20.0}_{-20.0-20.0}  $ &  $74.1^{+2.1+5.7}_{-3.3-5.1} $ & $75.6^{+2.4+6.4}_{-3.5-5.8} $ \\ 

  $\sigma_{8} $ & $0.940^{+0.120+0.200}_{-0.140-0.170}  $ &  $0.956^{+ 0.120+0.160}_{-0.070-0.190} $  & $ 0.875^{+ 0.022+0.049}_{-0.027-0.047}  $  & $ 0.875^{+0.021+0.055}_{-0.030-0.051}$ \\ 

$r_{\rm drag}$   & $ 147.23^{+ 0.29+0.59}_{-0.29-0.59}$ &  $ 147.09^{+ 0.30+0.54}_{-0.27-0.58}$    & $ 147.24^{+ 0.25+0.48}_{-0.25-0.48}$& $ 147.11^{+ 0.26+0.49}_{-0.26-0.52}$   \\ 
\hline

 
  \end{tabular}
  \caption{Constraints at 68\% and 95\% CL on free and some derived 
parameters under $w_0\mbox{-}w_a$CDM model baseline from the considered data combinations. 
The parameter $H_0$ is measured in the units of km/s/Mpc, $r_{\rm drag}$ in Mpc, whereas 
$M_{\nu}$ is in the units of eV.}
\label{CPL_results}
\end{table*}

\begin{table} 
\centering
\begin{tabular}{c|c|c}
\hline
\hline
            &  \,Planck + BAO + R19\, & \,Planck + BAO + R19\,  \\
\hline 
Parameter\, &  $w_0\mbox{-}w_a$CDM + $M_{\nu}$ & $w_0\mbox{-}w_a$CDM  \\
\hline 

$10^{2}\omega_{b }$ & $2.242^{+0.015+0.029}_{-0.015-0.030}$ & $2.245^{+0.014+0.028}_{-0.014-0.027}$ \\ 

$\omega_{_{\rm CDM }}  $ & $0.1194^{+0.0012+0.0024}_{-0.0012-0.0025}$ & $0.1192^{+0.0011+0.0023}_{-0.0011-0.0022}$	\\

$100 \theta_{*}$ & $1.04195^{+0.00030+0.00056}_{-0.00030-0.00058}$ & $1.04193^{+0.00029+0.00058}_{-0.00029-0.00056}		$\\

$\ln10^{10}A_{s }$ & $3.045^{+0.016+0.032}_{-0.016-0.030}	$ & $3.037^{+0.014+0.027}_{-0.014-0.028}$\\

$n_{s } $ & $0.9658^{+0.0043+0.0086}_{-0.0043-0.0083}$ & $0.9667^{+0.0041+0.0080}_{-0.0041-0.0084}$\\

$\tau_{reio }   $ & $0.0552^{+0.0081+0.017}_{-0.0081-0.015}$ & $0.0520^{+0.0074+0.015}_{-0.0074-0.015}$ \\

$w_0      $ & $-0.920^{+0.15+0.23}_{-0.095-0.27}	$ & $-0.89^{+0.17+0.25}_{-0.11-0.29}$\\

$w_a     $ & $-1.39^{+0.18+0.97}_{-0.59-0.64}$ & $-1.23^{+0.30+1.0}_{-0.68-0.80}$\\

$M_{\nu}$ & $< 0.31$ & ---\\

$\Omega_{m}    $ & $0.2602^{+0.0092+0.019}_{-0.0092-0.017}$ & $0.2594^{+0.0087+0.017}_{-0.0087-0.017}	$\\

$H_0$ & $74.2^{+1.4+2.6}_{-1.4-2.6}	$ & $73.9^{+1.2+2.4}_{-1.2-2.4}		$ \\

$\sigma_8$ & $0.864^{+0.015+0.030+}_{-0.015-0.029}$ & $0.874^{+0.014+0.028}_{-0.014-0.027}$ \\

$r_{\rm drag}        $ & $147.18^{+0.27+0.52}_{-0.27-0.53}	$ & $147.23^{+0.25+0.50}_{-0.25-0.48}$ \\
\hline

\end{tabular}
\caption{Constraints at 68\% and 95\% CL on free and some derived 
parameters under $w_0\mbox{-}w_a$CDM model baseline from the considered data combinations. 
The parameter $H_0$ is measured in the units of km/s/Mpc,, $r_{\rm drag}$ in Mpc, 
whereas $M_{\nu}$ is in the units of eV}
\label{CPL_results_2}
\end{table}

\section{Combined analyses of CMB plus transversal BAO data}
\label{results}

Throughout this section we will present our main results. 

\subsubsection{$\Lambda$CDM scenario}

For the $\Lambda$CDM + $M_{\nu}$ model, we summarize the main observational results in 
Table~\ref{LCDM_results}. 
For comparison, we also show analyses without and with neutrinos. 

\begin{figure*} 
\centering
\includegraphics[scale=0.6]{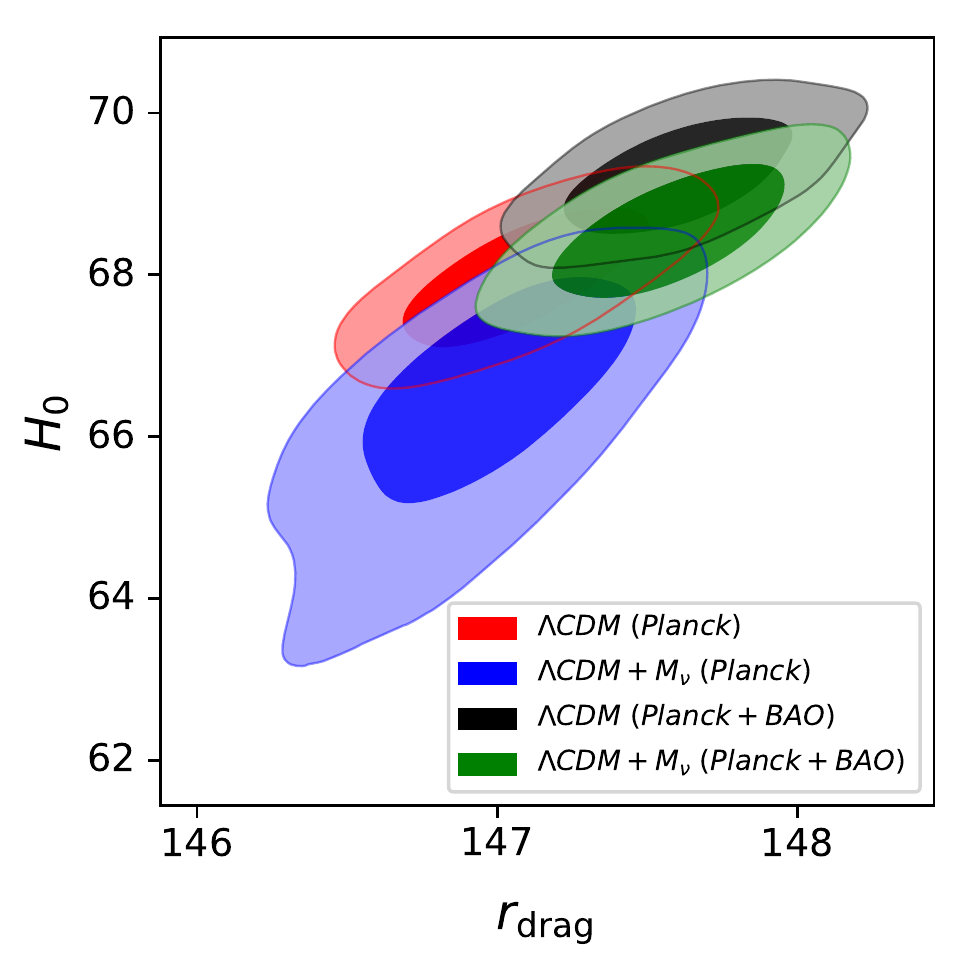} \,\,\,\,\,\,\,\,
\includegraphics[scale=0.6]{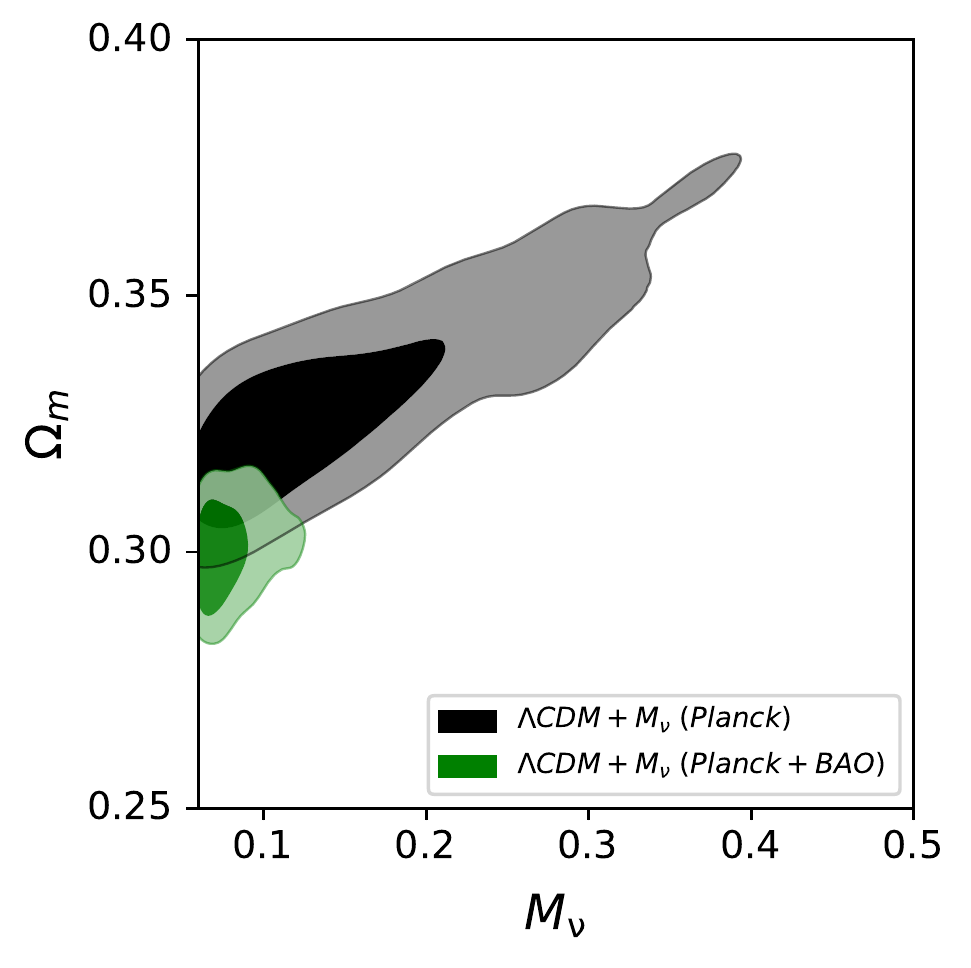}
\caption{Results from the $\Lambda$CDM cosmology. 
Left panel: Constraints at 68\% and 95\% CL in 
the parametric space $r_{\rm drag}$ - $H_0$ from Planck data only and Planck + transversal BAO, with and without the addition of neutrinos massa scale $M_{\nu}$ as a free parameter. Right panel: Constraints in the plan $\Omega_{m0}$ - $M_{\nu}$, under perspectives of the scenario $\Lambda$CDM + $M_{\nu}$.}
\label{fig:LCDM}
\end{figure*}

\begin{figure*} 
\centering
\includegraphics[scale=0.6]{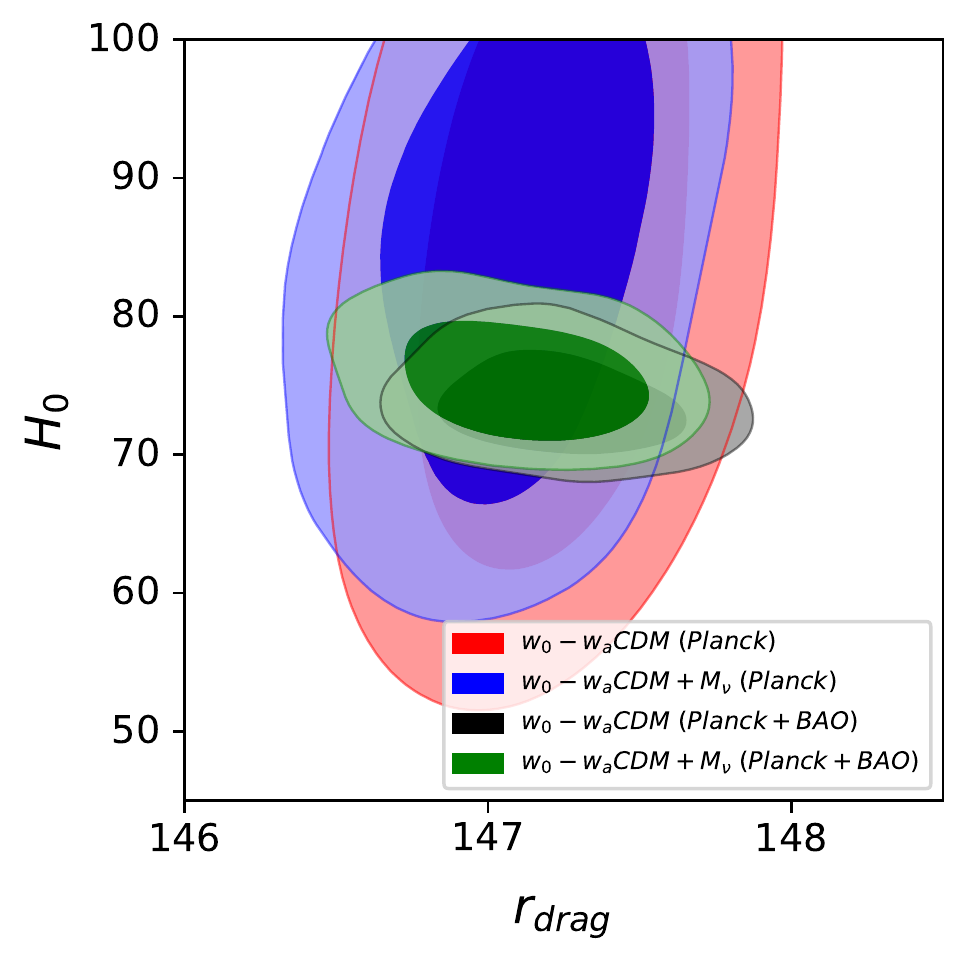} \,\,\,\,\,\,\,\,
\includegraphics[scale=0.6]{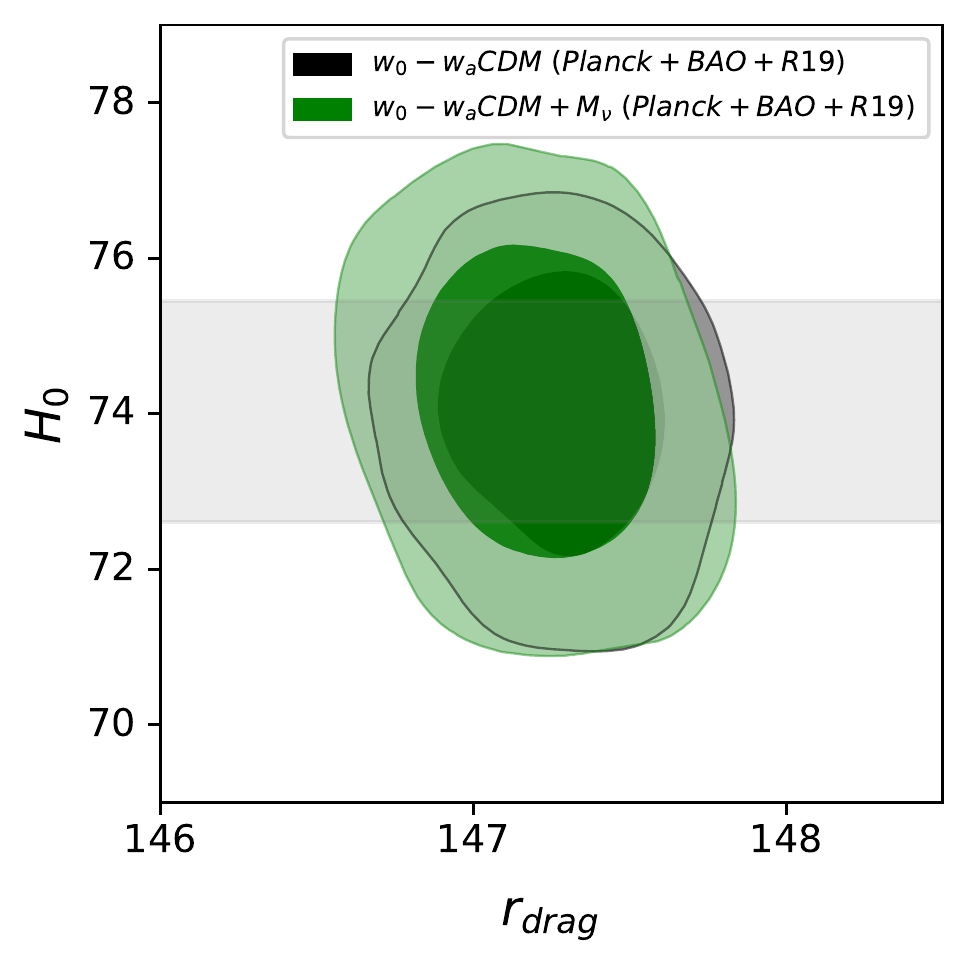}
\caption{Left panel: The 68\% CL. and 95\% CL. regions in the plan $r_{\rm drag}$ - $H_0$ inferring from $w_0\mbox{-}w_a$CDM model using Planck data only and Planck + transversal BAO. Right panel: The same as left panel, but a joint analysis from Planck + transversal BAO + R19. The vertical light red band corresponds to measure $H_0=74.03 \pm 1.42$ km s${}^{-1}$ 
Mpc${}^{-1}$.}
\label{fig:CPL_1}
\end{figure*}

\begin{figure*} 
\centering
\includegraphics[scale=0.6]{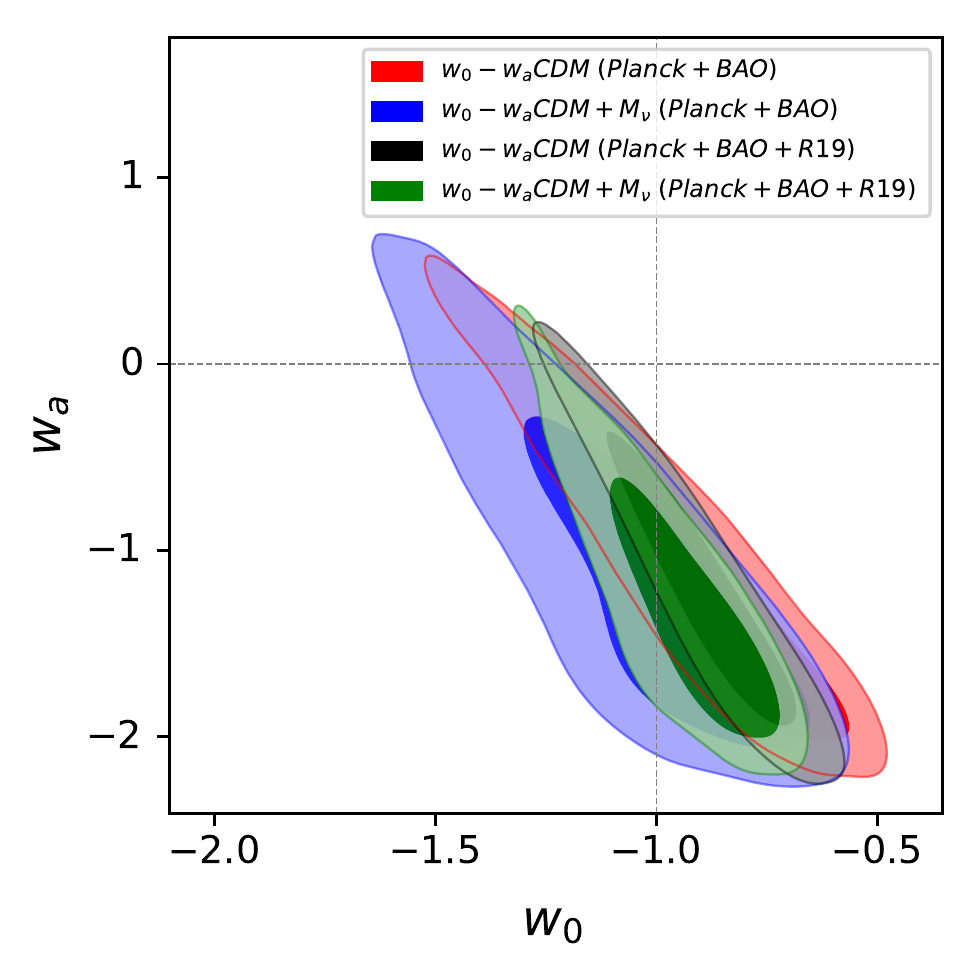} \,\,\,\,\,\,\,\,
\includegraphics[scale=0.6]{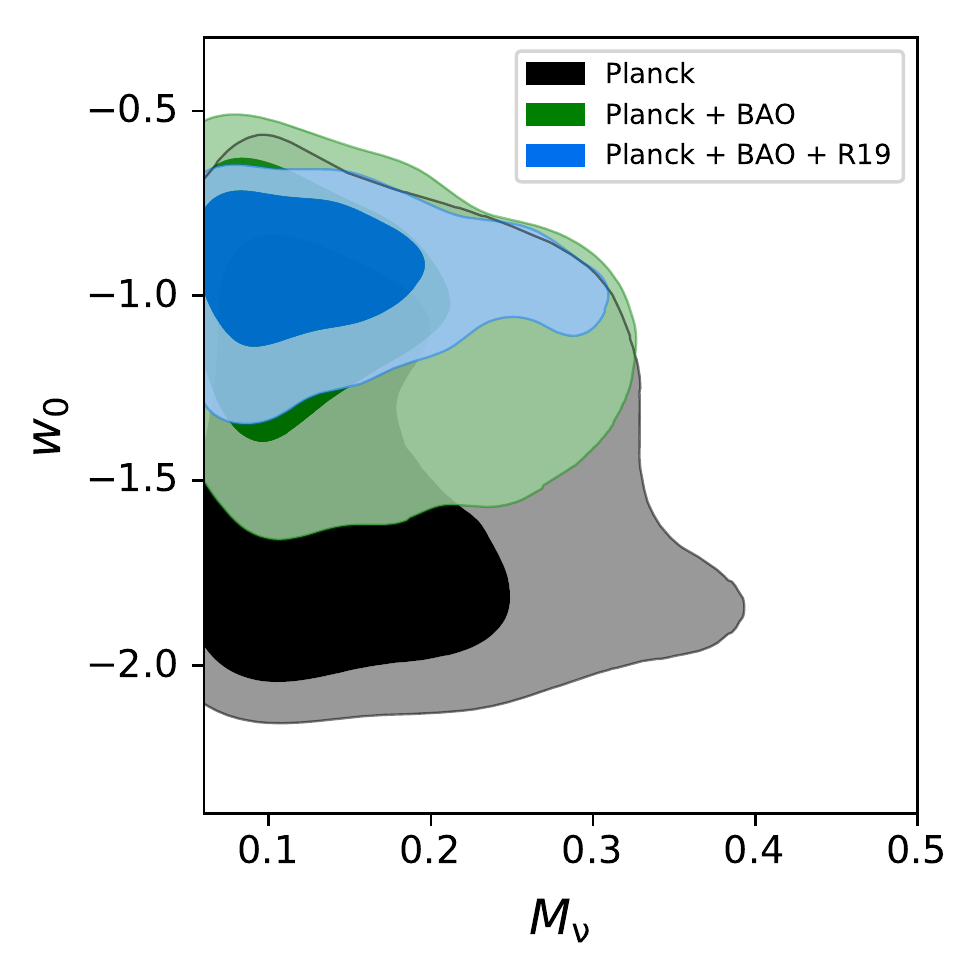}
\caption{Left panel: Parametric space at 68\% CL and 95\% CL in the plan $w_0 - w_a$ from the considered data combinations. Right panel: Confidence regions at 68\% CL and 95\% CL in the plan $M_{\nu} - w_0$ from the $w_0\mbox{-}w_a$CDM model in terms of the considered data combinations.}
\label{fig:CPL_2}
\end{figure*}

Assuming the $\Lambda$CDM scenario, we notice that adding the BAO data in a combined analysis with Planck, the constraints on the parameters that are 
most sensitive to geometrical tests, like $\Omega_{m}$ and $H_0$, are significantly improved. 
In fact, from Planck + transversal BAO analysis, the constraint on $H_0$ and $\Omega_{m}$ is significantly deviated to a higher and lower fit value, respectively, when compared with the analysis from Planck data only. 
However, the difference on these parameters between the analyses Planck + transversal BAO 
versus Planck data only, as well as for the other parameters in the baseline are compatible with each other at 68\% CL. Thus, we can note compatibility between Planck and transversal BAO data, including the fact that the 
addition of BAO data helps to break possible degeneracy in the parameters space, is an excellent outcome of the current analyses.

From the Planck data only, we obtain $H_0 = 67.39 \pm 0.56 $ km s${}^{-1}$ Mpc${}^{-1}$, while $H_0 = 69.23 \pm 0.50$ km s${}^{-1}$ Mpc${}^{-1}$ from Planck + transversal 
BAO data, in both cases at 68\% CL. 
The value obtained by the Planck team from CMB + BAO data\footnote{See section 5.1 
in~\cite{planck_VI} for details of the BAO data points 
used in the analyses done by the Planck collaboration.} is $H_0 = 67.77 \pm 0.42$. 
It is well reported in the literature that there is a strong tension between the $H_0$ estimates 
done by the Planck collaboration and local measurements as reported by Riess et 
al.~\cite{Riess2019}, i.e., $H_0 = 74.03 \pm 1.42$ km s${}^{-1}$ Mpc${}^{-1}$. 
One can notice that the combined analyses done here, using Planck + transversal BAO data, 
minimally alleviate this tension, but a tension at more than $3\sigma$ still remains. The difference on the $H_0$ parameter between the Planck team constraints and our results is approximately $\sim 1.4\sigma$.

In the left panel of Figure~\ref{fig:LCDM}, we show the confidence level contours in the parametric space $r_{\rm drag}$ - $H_0$ from all our analyses using $\Lambda$CDM model. Some words are in due here regarding $r_{\rm drag}$. This quantity is not directly measured by CMB data, its calculation depends on model hypotheses of early time physics, for this it is obtained in a model-dependent way in CMB analyses~\citep{Sutherland,Heavens,planck_VI}. Both parameters, $r_{\rm drag}$ and $H_0$, provide an absolute scale for distance measurements at opposite ends of the observable universe, $r_{\rm drag}$ (early time) and $H_0$ (late time). 
When measured with the same data, these parameters must agree with the values predicted by the standard cosmological model. Otherwise, significant deviations of these parameters with respect to the expected values would provide indications for some new physics beyond the standard model, or unaccounted systematic errors in the measurements. Planck team reported the value $r_{\rm drag} = 147.21 \pm  0.23$ from CMB + BAO at 68\% CL in $\Lambda$CDM cosmology~\citep{planck_VI}. 
These estimates are not only compatible, but very similar to ours, in all analyses (see Table~\ref{LCDM_results}). On the other hand, a model-independent reconstruction
from the early-time physics shows that $r_{\rm drag} = 136.7 \pm  4.1$~\citep{Bernal}. As argued by~\citet{Bernal}, this strong tension in the $r_{\rm drag}$ measurement is entirely due to the tension in the $H_0$ parameter, via the strong correlation between $H_0$ and $r_{\rm drag}$ parameters.  Considering a physics beyond the standard model, in ref.~\citep{Kumar_Nunes_Yadav} it is assumed a dark coupling between dark matter and photons, an approach that can reconcile the tension in both parameters, $H_0$ and $r_{\rm drag}$. Another proposal for some new physics in light the $H_0$ and $r_{\rm drag}$ tension at early and/or late time modification in the standard cosmological model was also proposed in ~\citep{T01,T02,T03,M07,T05,T06,T07,T08}. 

In the right panel of Figure~\ref{fig:LCDM}, we show the confidence level contours 
in the parametric plane $\Omega_{m}$ - $M_{\nu}$, where we find $M_{\nu} < 0.11$ eV at 95\% CL from Planck + transversal BAO data. That boundary on the neutrino mass scale is practically the same as that one reported by the Planck team using Planck CMB + BAO, i.e., $M_{\nu} < 0.12$~eV. Some minimal displacement can be noted on the $\Omega_{m}$ best fit value, but again, the constraints are fully compatible with each other at 68\% CL. Thus, we conclude that the combination of the transversal BAO and CMB data can bound $M_{\nu}$ with the same accuracy than other joint analyses reported in the literature. 
\\

\subsubsection{$w_0\mbox{-}w_a$CDM scenario}

In view of the capacity of the transversal BAO data to breaks the degeneracy on some cosmological parameters, let us study how these data could bound some dynamical effect of dark energy. 
In Table~\ref{CPL_results}, we summarize the results of our statistical analyses from the perspective of the $w_0\mbox{-}w_a$CDM model. As already known, assuming a $w_0\mbox{-}w_a$CDM model, the constraints on $H_0$ become degenerate in such way to obtain high $H_0$ values enough to be compatible with local measurements.. We combine Planck data with transversal BAO data to break the degeneracy on the full baseline parameters of the model, finding $H_0 = 74.1^{+2.1}_{-3.3}$ km s${}^{-1}$ Mpc${}^{-1}$ (without neutrinos) and $H_0 = 75.6^{+2.4}_{-3.5}$ km s${}^{-1}$ Mpc${}^{-1}$ (including neutrinos), both at 68\% CL. Thus, we can clearly notice that by adding transversal BAO  data, the analysis significantly improves the bounds on $H_0$ when compared to Planck  data only. More specifically, it is obtained an  improvement of $\sim$24\% and $\sim$32\% on $H_0$ in the analysis without and with neutrinos, respectively. See $H_0$ value from Planck data only in Table~\ref{CPL_results} for this same scenario. Additionally, these constraints are fully compatible with local estimate of $H_0$ from HST.
Thus, the current tension on $H_0$ present in $\Lambda$CDM model, does not persist within this scenario, and the combination Planck + transversal BAO is not in tension with local measures. We conclude that dynamical models, like $w_0-w_a$CDM, can solve the tension on the $H_0$ parameter. In view of this, let us also consider the joint analysis Planck + transversal BAO + R19 data. These results are summarized in Table~\ref{CPL_results_2}. 

Figure~\ref{fig:CPL_1} shows the parametric space in the plane $r_{\rm drag}$ -- $H_0$. On the left panel, we quantify the improvements due to the inclusion of the transversal BAO data. On the right panel, we have the joint analyses Planck + BAO + R19. With respect to the $r_{\rm drag}$ parameter, we do not notice any significant deviations as compared with the results predicted for the minimum $\Lambda$CDM model. Figure~\ref{fig:CPL_2} on the left panel shows the constraints in the plane $w_0 -w_a$ from Planck + BAO and Planck + BAO + R19 data combination. We can see that transversal BAO data set significantly improves the 
constraints on the EoS parameters, in comparison with Planck data only. We find an improvement of $\sim$54\% and $\sim$38\% on $w_0$ and $w_a$, respectively, from the analysis without the presence of neutrinos. 
Instead, with neutrinos we find an improvement of $\sim$67\% and $\sim$39\% on $w_0$ and $w_a$, respectively. 
We found no evidence for deviations of the minimum $\Lambda$CDM cosmology, even when 
including the R19 datum in the analysis. 
In the right panel of Figure \ref{fig:CPL_2} we show the relationship between $w_0 - M_{\nu}$ and 
the effect on the confidence contour levels due to diverse combined data analyses. 
We observe that the bound on the neutrino mass scale is slightly decreased, while the constraints 
on the $w_0$ parameter are more robust, significantly improving the restrictions when considering 
the joint analysis Planck + BAO and Planck + BAO + R19. 
We found $M_{\nu} < 0.38, \,\, 0.33, \,\, 0.31$ eV at 95\% CL, from Planck,  Planck + BAO, 
and Planck + BAO + R19 data sets, respectively. 
The final effect is that the presence of a dynamical dark energy component slightly extends the bound on $M_{\nu}$, as compared to $\Lambda$CDM model prediction. 
Effects of the neutrino mass scale on some dynamical dark energy models are also 
discussed in~\citep{neutrinos_DE1,neutrinos_DE2,neutrinos_DE3}, but these studies consider 
other data sets. Our results represents a new update on $M_{\nu}$ through the use of our recent transversal BAO data compilation. 

The fact that $w_0 \mbox {-} w_a $CDM scenario can generate high $H_0$ values, is due the feature that the scenario predict less dark matter today --in  contrast, more dark energy-- via the relation $\Omega_{m} + \Omega_{DE} = 1$, where 
$\Omega_{m} = \Omega_{b} + \Omega_{DM}$, in direct comparison with the $\Lambda$CDM  best fit values. 
Notice that $\Omega_{b}$ is fully compatible in all these scenarios. So, the change on $\Omega_{m}$ estimates is due to dark matter density only, once the radiation 
(photons + neutrinos) contribution is negligible at $z =0$. 
Because this scenario predicts more dark energy at late times, the universe expands faster 
than predicted in the $\Lambda$CDM cosmology, generating a larger $H(z)$ and, 
at the same time, changing the slope of the Sachs-Wolfe plateau, that is, the late-time integrated 
Sachs-Wolfe effect (ISW), where the amplitude of the ISW effect will depend on the duration of the 
dark energy-dominated phase, which is basically managed by the ratio 
$\Omega_{DM}/\Omega_{DE}$. 
The $H_0$ value from CMB data is inferred analyzing the first acoustic peak position, which 
depends on the angular scale $\theta_{*} = d_s^{*}/D_A^{*}$, where $d_s^{*}$ is the sound horizon 
at decoupling (the distance a sound wave traveled from the big bang to the epoch of the 
CMB-baryons decoupling) and $D_A^{*}$ is the angular diameter distance at decoupling, which 
in turn depends on the expansion history, $H(z)$, after decoupling, controlled also by the ratio 
$\Omega_{DM}/\Omega_{DE}$ and $H_0$ mainly. 
The $w_0\mbox{-}w_a$CDM scenario is changing primarily the $D_A^{*}$ 
history, because a faster expansion at late times increases the 
angular diameter distance to the surface of last scattering, thus 
generating high estimates on $H_0$ parameter.

\section{Model-independent reconstruction of the deceleration parameter}
\label{qz_section}

In this section, we will derive model-independent constraints on 
the deceleration parameter, $q(z)$, directly from analyses of the transversal BAO data. In order to do so, we use the so-called Gaussian Processes (GP)  \citep{seikel2012,Rasmussen}. The GP has been shown to be a powerful tool to investigate cosmological parameter to some model-independent way \citep{GP1,GP2,GP3,GP4,GP5}. In what follows, we briefly describe the methodology.

The GP method consists of considering Gaussian errors on data, so that the function that should describe the data correctly could be seen as a random normal variable. The method is explained in  refs.~\cite{Rasmussen,seikel2012,ztGP}. As the data points are expected to be related through the same underlying function $f(x)$, two points $x$ and $x'$ are correlated through a covariance function (or kernel) $k(x,x')$. By choosing such a covariance function, the distribution of  functions is described by
\begin{align}
\mu(x)&=\langle f(x)\rangle,\nonumber\\ k(x,x')&=\langle(f(x)-\mu(x))(f(x')-\mu(x'))\rangle,\nonumber\\
\mathrm{Var}(x)&=k(x,x).
\label{moments}
\end{align}

There are many choice options of the covariance functions, but without loss of generality, we shall focus on the Gaussian (or Squared Exponential) kernel, which is given by
\begin{equation}
k(x,x')=\sigma_f^2\exp\left[-\frac{(x-x')^2}{2l^2}\right],
\end{equation}
where $\sigma_f$ and $l$ are the so called hyperparameters. The GP method consists on optimizing for $\sigma_f$ and $l$ and then using \eqref{moments} for reconstruct the function $f(x)$.

We use the freely available software GaPP\footnote{\url{http://www.acgc.uct.ac.za/~seikel/GAPP/index.html}} in order to reconstruct $q(z)$ from the $\theta_\text{BAO}$ data. 
%
%
First, we have tried to reconstruct $\theta_\text{BAO}(z)$. However, we have found that the reconstruction of $\theta_\text{BAO}(z)$ does not yield reliable results. As explained in \cite{seikel2012}, given the same amount of data, functions that change very rapidly are more difficult to reconstruct than smooth functions. It happens that $\theta_\text{BAO}(z)$ is not an smooth function of the redshift $z$. 
In fact, for any cosmological model, $D_A(z=0)=0$. As  $\theta_\text{BAO}(z)\propto 1/D_A$, then $\theta_\text{BAO}(z)\rightarrow\infty$ for $z\rightarrow0$.

%
Instead, we  reconstruct $D_A(z)$, 
which is expected to be a smooth function of the redshift ($D_A\propto z$ at low redshift for any cosmological model). In order to obtain the $D_A(z)$ data, we have used Eq. \eqref{eq_bao} to obtain the uncertainties through error propagation as 
\begin{equation}
 \sigma_{D_A}=D_A\frac{\sigma_{\theta_\text{BAO}}}{\theta_\text{BAO}} \, .
\end{equation}

\begin{figure*}[ht!]
 \centering
 \includegraphics[scale=0.5]{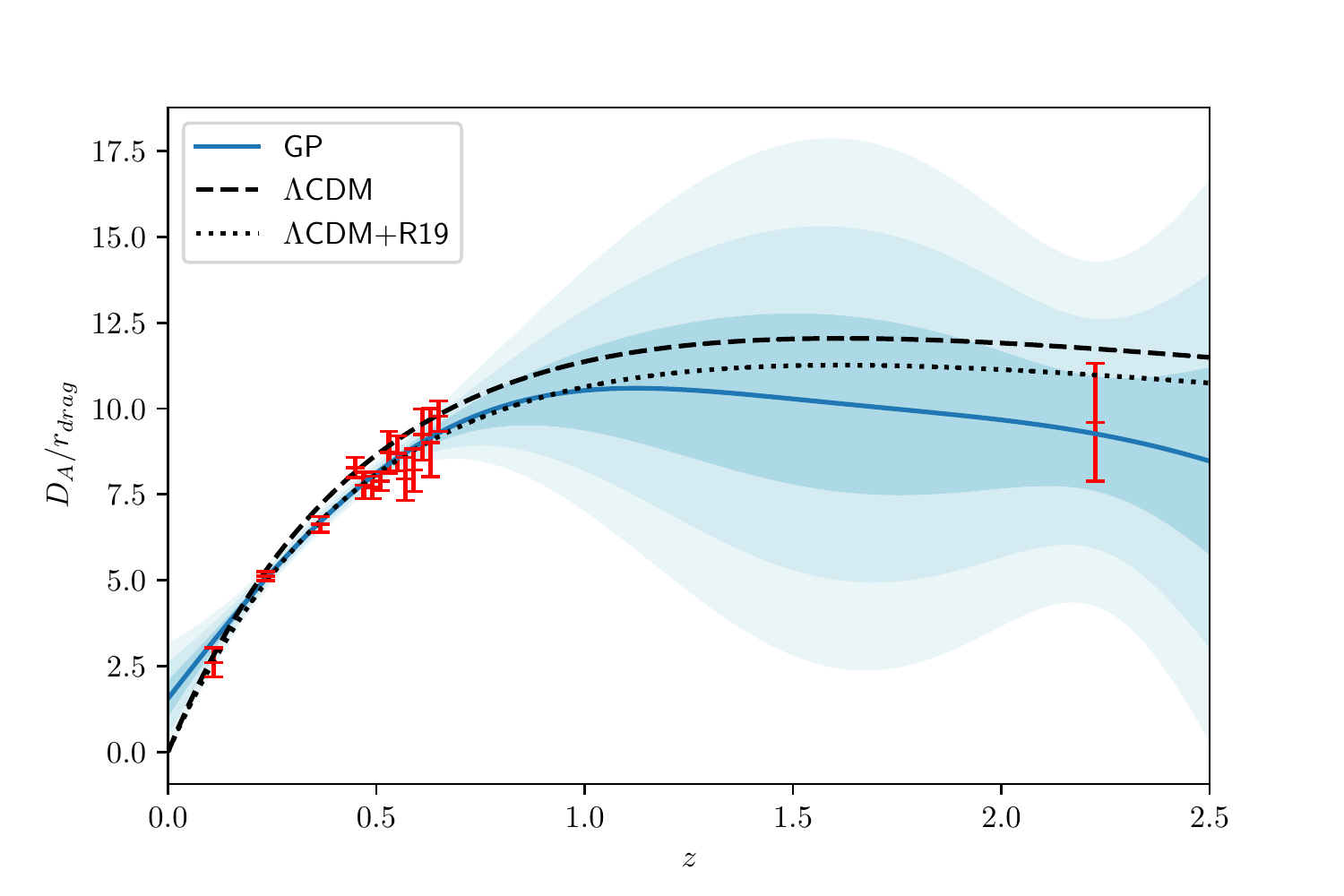}
 \includegraphics[scale=0.5]{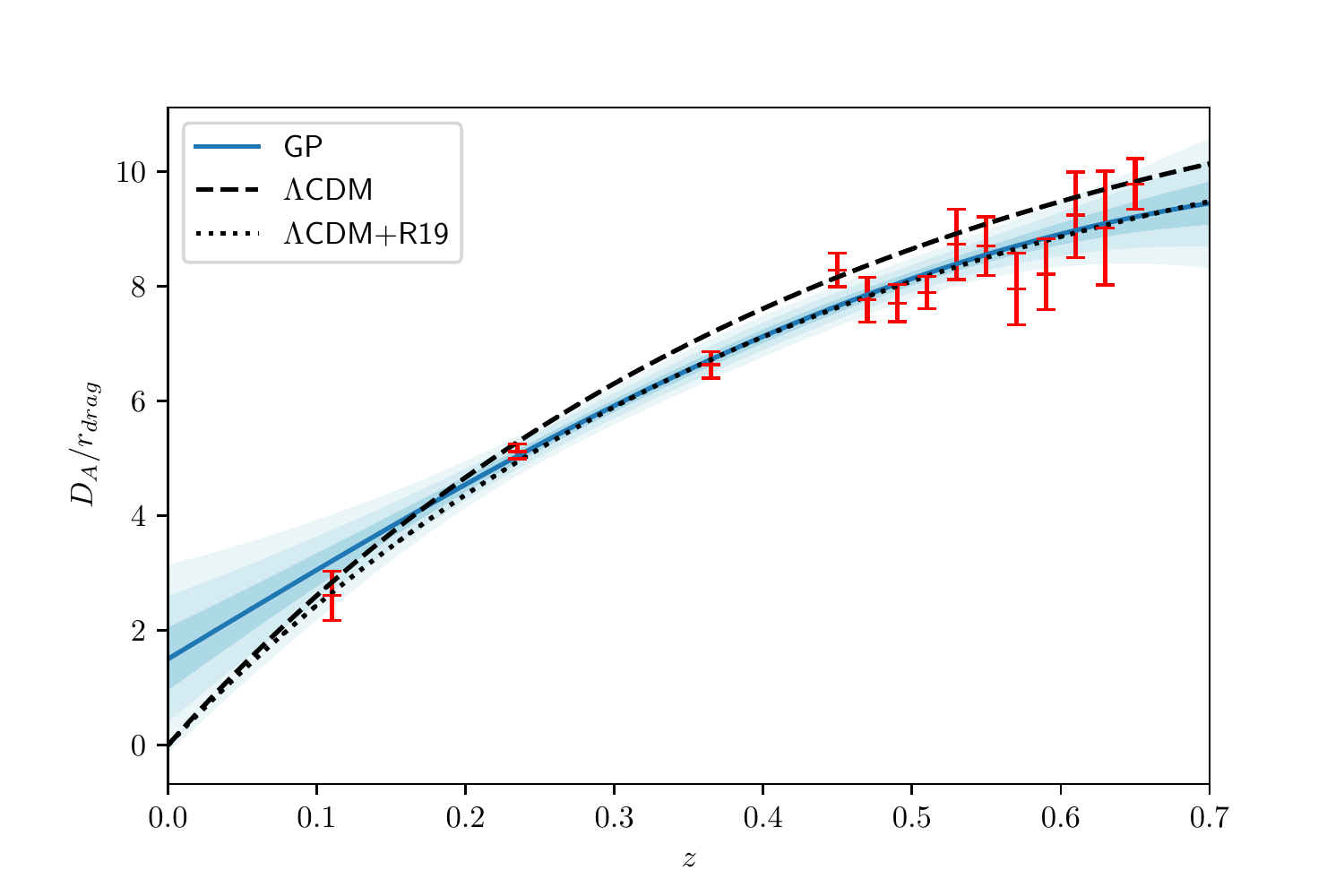}
 \caption{Reconstruction of $D_A(z)/r_\text{drag}$. The blue solid line corresponds to the median of $D_A/r_\text{drag}$. Also shown are the 68.3\%, 95.4\% and 99.7\% CL regions. Left: Full sample. Right: Sample with $z<0.7$.}
 \label{DA}
\end{figure*}

\begin{figure*}[h!]
 \centering
 \includegraphics[scale=0.5]{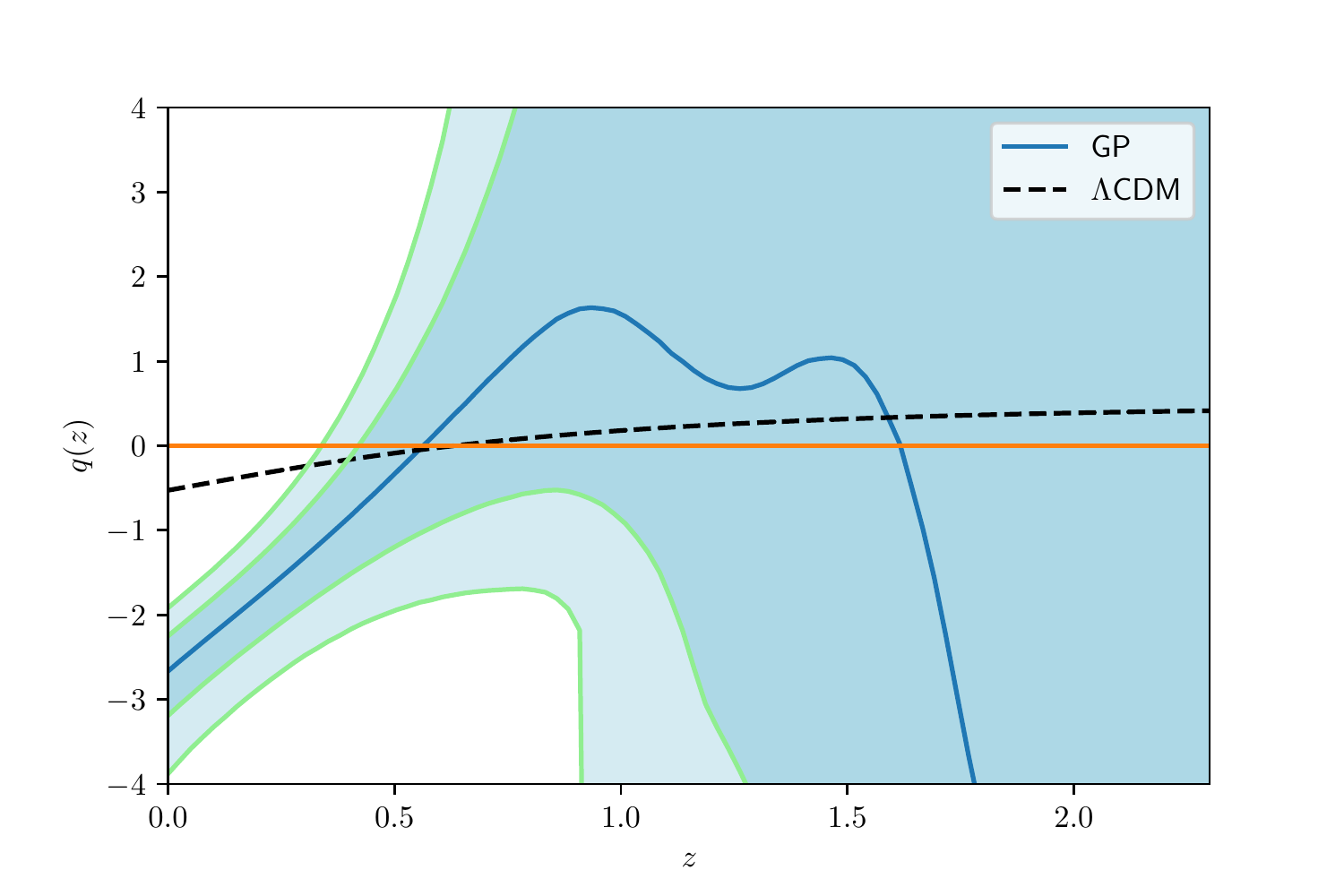}
 \includegraphics[scale=0.5]{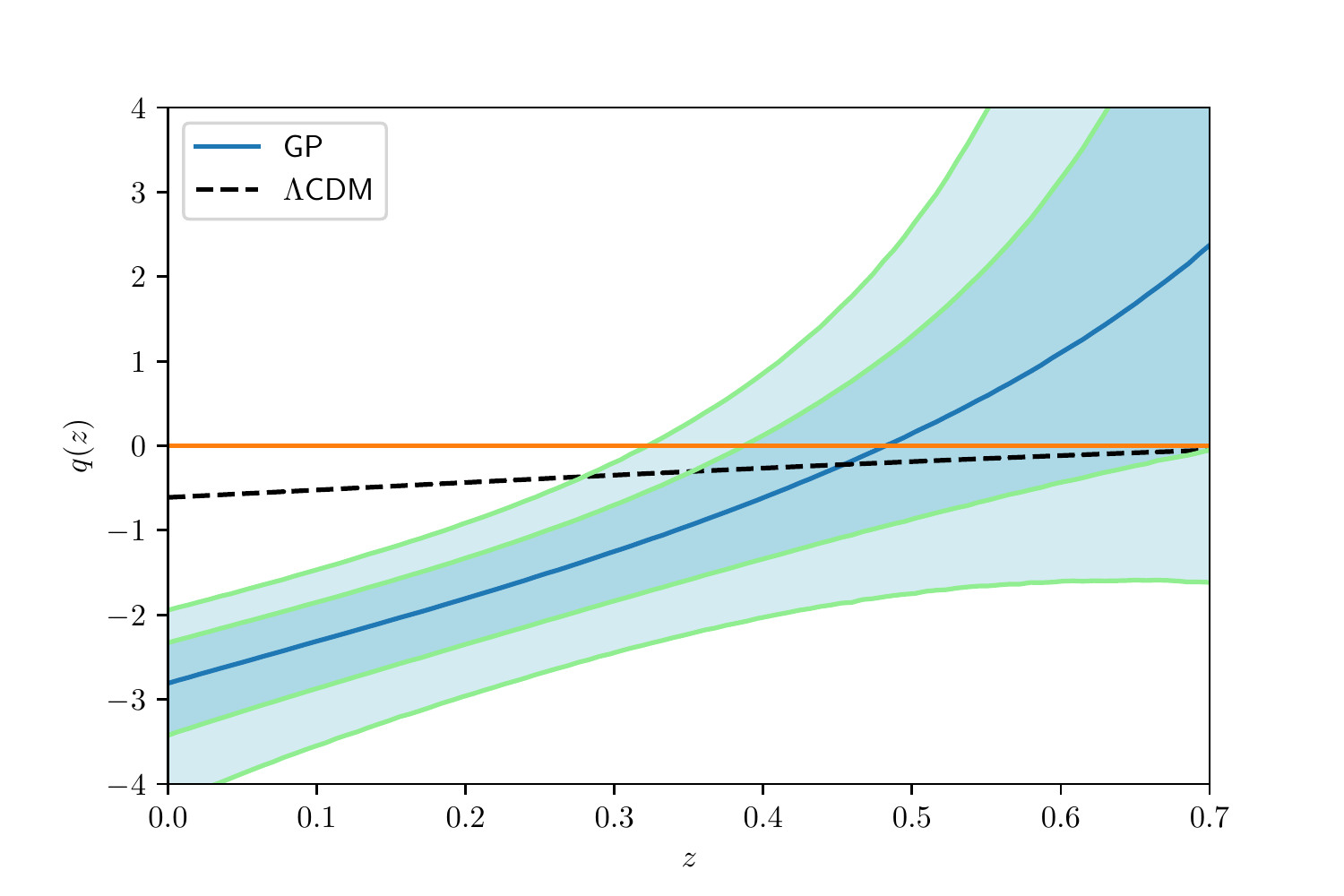}
 \caption{Reconstruction of $q(z)$. Also shown are the 68\% and 95\% CL regions. Left: Analyses of the full BAO data sample. Right: Analyses considering the sub-sample with $z<0.7$.}
 \label{qzrec}
\end{figure*}

The $D_A(z)$ reconstruction can be seen on Figure~\ref{DA}. As observed in this figure (left panel), the reconstruction performs poorly in the range $0.7 \lesssim z \lesssim 2$, due to lack of data on this interval. Aiming to focus on the region with more available data, we perform a second reconstruction with BAO data only up to $z<0.7$. 
This is shown in Figure~\ref{DA} (right panel).

As can be seen in Figure~\ref{DA} (right panel), $D_A/r_\text{drag}$ is better constrained on $0.4<z<0.7$, where there is more data. For $z<0.4$, there are only 3 data points, with a larger uncertainty for the lowest redshift datum. We believe this is the reason why the reconstruction poorly represents the expected behavior at low redshift, yielding a distance compatible with zero at $z=0$ only at the 99.7\% CL.

Also observed in Figure~\ref{DA}, two predictions from $\Lambda$CDM,  both using $\Omega_m=0.292$ and $r_\text{drag}=147.59$ Mpc, from the best fit of Planck+BAO combination. However, we have found that $D_A$ is much sensitive to the choice of $H_0$. Thus, we show a curve with $H_0=69.23$ km s${}^{-1}$ Mpc${}^{-1}$ from Planck + BAO (dashed curve) and one with $H_0=74.03$ km s${}^{-1}$ Mpc${}^{-1}$ from R19 data (dotted curve). As can be seen the curve with higher value of the Hubble constant yields a better agreement with the data and with the GP  reconstruction.

Assuming spatial flatness and the Etherington duality relation, $D_L(z)=(1+z)^2D_A(z)$, and following the methodology given in ref.~\cite{ztGP} (see section 3), we can write $q(z)$ in the form
\begin{equation}
q(z)=-\frac{(1+z)^2D_A''(z)+3(1+z)D_A'(z)+D_A(z)}{D_A(z)+(1+z)D_A'(z)} \, .
\end{equation}

From this result, the $q(z)$ function is independent of the distance  dimension, so it is independent of $r_\text{drag}$. We show the result of this reconstruction on Fig.~\ref{qzrec}. The blue solid line corresponds to the median $q(z)$ obtained from the GP. The light blue regions correspond to the 68\% and 95\% CL around the median. These regions were found by sampling the multivariate normal distribution of $D_A$, $D_A'$, and $D_A''$ found from the reconstruction above. We also show the theoretical prediction for $\Lambda$CDM (dashed curve) 
assuming $\Omega_m=0.292$ from Planck+BAO constraint. As can be seen, the $\Lambda$CDM model presents some tension with the reconstructed function done in light of the transversal BAO data only at low redshift, $z<0.3$. 
This is probably due to the lack of data on this $z$ range. On the other hand, in the interval $0.3<z<0.7$, with more data, there is a nice agreement with the $\Lambda$CDM cosmology using  this model-independent reconstruction.

\section{Final Remarks}
\label{final}

We have presented a new compilation with 15 angular BAO measurements summarized in Table~\ref{table1} obtained from analyses of luminous red galaxies, blue galaxies, and quasars 
catalogs using  various public data releases from the SDSS collaboration. 
As explained in the refs.~\citet{Sanchez11,Carnero,Gabriela1}, these transversal BAO data are weakly dependent on a cosmological model. 
It is worth mentioning that, due to the cosmological-independent methodology used to perform these transversal BAO measurements the errors calculated are larger than the errors obtained using a fiducial cosmology to model the BAO signal. 
The reason for this fact is that, while in the former methodology the error is given by the 
measure of how large is the BAO bump, in the later approach the model-dependent best-fit 
of the BAO signal determines a smaller error. 
Typically, in the former methodology the error can be of the order of $\sim 10\%$, but in 
some cases it can arrive to 18\%, instead, in the model-dependent approach it is of the order of few percent~\citep{Sanchez11}.

For the first time it is performed a combined analyses using these 
BAO data compilation, to explore the parameter space of some DE models and, additionally, obtain a precise estimate of 
$r_{\rm drag}$ from each model. 
Furthermore, we derive new bounds on the neutrino mass scale 
$M_{\nu}$ within $\Lambda$CDM and extended models, considering 
the possibility for some dynamical DE scenarios. 
In addition, we perform an independent model analyses to 
reconstruct the deceleration parameter $q(z)$ from the compilation 
of these BAO data. 

An interesting outcome of the current analyses is that the 
addition of transversal BAO data helped to break possible degeneracy in the parameters space, showing the compatibility between Planck and 
these transversal BAO data sets.  Last, but not least, we mention that the combination of the transversal BAO and CMB data can bound $M_{\nu}$ with the same accuracy than other joint analyses reported in the literature. Finally, we 
observe that in the $w_0-w_a$CDM model one can solve the  current tension on the $H_0$ parameter. Our results show that this current compilation with 15 BAO measurements,  obtained free of any fiducial cosmology, can generate observational constrains compatible with other BAO compilations present in the literature. Therefore, it would be interesting to investigate these data in other cosmological contexts. For readers interested in using our BAO likelihood (python language), please contact us.

\section*{Acknowledgements}

\noindent
The authors thank the referee for some clarifying points.
RCN would like to thank the Brazilian Agency FAPESP for financial  support under Project
No. 2018/18036-5 and thanks the hospitality of Observat\'orio Nacional, where part of this work was carried out. AB acknowledges a CNPq fellowship. 

\section*{Data availability}
The data underlying this article will be shared on request to the corresponding author.

\end{document}